\documentstyle[amssymb,aps,psfig]{revtex}

\begin{document}
\draft
\title{Microcanonical Thermostatistical Investigation\\
of the Blackbody Radiation}
\author{L. Velazquez$^{a,}$\thanks{%
luisberis@geo.upr.edu.cu}and F. Guzm\'{a}n$^{b,}$\thanks{%
guzman@info.isctn.edu.cu}}
\address{$^{a}$Departamento de F\'{i}sica, Universidad de Pinar del R\'{i}o\\
Mart\'{i} 270, esq. 27 de Noviembre, Pinar del R\'{i}o, Cuba. \\
$^{b}$Departamento de F\'{i}sica Nuclear\\
Instituto Superior de Ciencias y Tecnolog\'{i}as Nucleares\\
Quinta de los Molinos. Ave Carlos III y Luaces, Plaza\\
Ciudad de La Habana, Cuba.}
\date{\today}
\maketitle

\begin{abstract}
In this work is presented the microcanonical analysis of the blackbody
radiation. In our model the electromagnetic radiation is confined in an
isolated container with volume $V$ in which the radiation can not escape,
conserving this way its total energy, $E$. Our goal is to precise the
meaning of the Thermodynamic Limit for this system as well as the
description of the nonextensive effects of the generalized Planck%
\'{}%
s formula for the spectral density of energy. Our analysis shows the
sterility of the intents of finding nonextensive effects in normal
conditions, the traditional description of the blackbody radiation is
extraordinarily exact. The nonextensive effects only appear in the low
temperature region, however, they are extremely difficult to detect. 
\end{abstract}


\section{Introduction}

In the present paper we recover the famous problem of the macroscopic
description of the blackbody radiation. In the last years some investigators
have been reconsidered some aspects of this theory due to the coming of new
ideas and conceptions during the development of a new thermodynamic: that
which allows the study of the nonextensive systems.

In this frame, this problem have been approached by mean of the very well
known {\em nonextensive thermodynamic} of Tsallis\cite{Buy,Mart,Tir1,Tir2}.
In general way this conception constitutes a generalization of the
Boltzmann-Gibbs formalism. It is based mainly in the introduction of
nonextensive entropy, $S_{q}$, which is a generalization of the
Shannon-Boltzmann-Gibbs extensive entropy\cite{Tsal1,Tsal2}:

\begin{equation}
S_{q}=-%
\mathrel{\mathop{\sum }\limits_{k}}%
p_{k}^{q}\ln _{q}p_{k}
\end{equation}
where $\ln _{q}x$ is the q-generalization of the logarithmic function:

\begin{equation}
\ln _{q}x\equiv \frac{x^{1-q}-1}{1-q}
\end{equation}

Among the fundamental aspects of this new conception we find the possibility
to deal the study of systems with potential distributions, that is, systems
with fractal characteristics. In spite of the attractiveness of this
formulation, the same one suffers of a logical internal inconsistence due to
dependence in all the theory of a phantasmagorical parameter, the entropic
index, $q$. This parameter represents the measure of the degree of
nonextensivity of the systems, an intrinsic characteristic of the same,
however it can not be inferred from the general principles. In fact, it acts
as a adjustment parameter of the theory, we need the experiment or the
computational simulation in order to precise it.

In the present system there are no long-range correlations due to the
presence of long-range interactions. In fact, this is a typical extensive
system.\ The only possible nonextensive effects in the theory of the
blackbody radiation could be found due to the non realization of the
Thermodynamic Limit, that is, due to the finite nature of the system.

It is very interesting to specify the meaning of the thermodynamic limit for
this system, as well as which are the main peculiarities of the physical
quantities product to the finite nature of the same one.

To reach these objectives we will make use of the main results of the {\em %
microcanonical thermostatistic} of D.H.E. Gross\cite{gro1,gro2}. This theory
returns to the pregibbsian times assuming the Boltzmann%
\'{}%
s definition of entropy:

\begin{equation}
S_{B}=\ln W\left( E;a\right)  \label{bol}
\end{equation}
his famous gravestone%
\'{}%
s epitaph. The thermodynamic formalism of this formulation has been
conceived in order to be equivalent in the thermodynamic limit with the
formalism of the classical thermodynamic for those systems that become in
this limit in a ordinary extensive systems\cite{gro3,gro4,gro5}, so that, it
could be applied to this kind of finite systems. Our problem is in fact a
typical example for the application of this theory.

\section{Microcanonical Analysis of the blackbody radiation.}

In the present section we will perform the microcanonical analysis of the
blackbody radiation. Firstly, we will expose some general aspects of the
microcanonical approaching of the quantum systems of identical
noninteracting particles. After, we will begin the analysis remembering some
results of the traditional thermodynamic. Subsequently the density of
accessible states will be calculated, being determined this way the Boltzmann%
\'{}%
s entropy of the system. It will be carried out a comparison between the
microcanonical temperature derived of this entropy with the absolute
temperature of the traditional description. This analysis will help to
specify the meaning of the thermodynamic limit for the blackbody radiation.

The generalization of the Planck%
\'{}%
s formula for the spectral density of energy will be obtained analyzing its
more significant particularities due to the finite nature of the system as
well as their behavior during the step to the thermodynamic limit.

\subsection{Microcanonical description of a quantum system of identical
noninteracting particles.}

In analogy to the classical case, the macroscopic state of the system is
determined by the knowledge of certain set of integrals of movement, $%
\left\{ I\right\} $. From the general theory it is known that macroscopic
observables are obtained by the relation:

\begin{equation}
\overline{O}=\frac{Sp\left( \widehat{O}\widehat{\rho }\right) }{Sp\left( 
\widehat{\rho }\right) }  \label{do}
\end{equation}
where $\widehat{\rho }$ is matriz density operator. For the microcanonical
description:

\begin{equation}
\widehat{\rho }_{M}\left( I;a\right) \equiv \delta \left( I-\widehat{I}%
\left( a\right) \right)
\end{equation}
where $a$ represent the external parameters of the systems. Let us pay
attention to the accessible states density of the system:

\begin{equation}
\Omega \left( I;a\right) =Sp\left( \delta \left( I-\widehat{I}\left(
a\right) \right) \right)
\end{equation}

For the special case of the systems of noninteracting particles is
convenient to make use of the integral representation of the Dirac delta
function:

\begin{equation}
\Omega \left( I;a\right) =\int d^{n}\widetilde{\xi }\exp \left( i\overline{%
\xi }\cdot I\right) Sp\left( \exp \left( -i\overline{\xi }\cdot \widehat{I}%
\left( a\right) \right) \right)  \label{de1}
\end{equation}
where $d^{n}\widetilde{\xi }=\frac{d^{n}\xi }{\left( 2\pi \right) ^{n}}$ and 
$\overline{\xi }=\xi -i\beta $ with $\beta \in {\bf R}^{n}$. We recognize
immediately the partition function of the canonical description, this time
with complex argument:

\begin{equation}
{\cal Z}\left( z;a\right) =Sp\left( \exp \left( -z\cdot \widehat{I}\left(
a\right) \right) \right) \text{ (}z=\beta +i\xi \text{)}
\end{equation}
Introducing the {\em Planck%
\'{}%
s potential}, ${\cal P}\left( z;a\right) $:

\begin{equation}
{\cal P}\left( z;a\right) =-\ln {\cal Z}\left( z;a\right)
\end{equation}
the Eq.[\ref{de1}] is rewritten as:

\begin{equation}
\Omega \left( I;a\right) =\int d^{n}\widetilde{\xi }\exp \left( i\overline{%
\xi }\cdot I-{\cal P}\left( i\overline{\xi };a\right) \right)  \label{id}
\end{equation}

It is very well known that the Planck%
\'{}%
s potential of a system of identical noninteracting particles is expressed
as:

\begin{equation}
{\cal P}\left( z;a\right) =-\eta 
\mathrel{\mathop{\sum }\limits_{k}}%
g_{k}\left( a\right) \ln \left[ 1+\eta \exp \left( -z\cdot i_{k}\left(
a\right) \right) \right]
\end{equation}
where the index $k$ represent the state with eingenvalues $i_{k}\left(
a\right) $ for the integrals of movement and degeneracy $g_{k}\left(
a\right) $. The constant $\eta $ is equal to $-1$ for the bosons or $+1$ for
fermions. When the thermodynamic limit takes place in the system the
microcanonical and the canonical description are equivalent if under the 
{\em scaling transformation} the integrals of movement and the Planck%
\'{}%
s potential are scaled homogeneously:

\begin{equation}
I\rightarrow \alpha I;\text{ }a\rightarrow \alpha ^{s}a\text{ }\left(
s=0,1\right) \Rightarrow {\cal P}\left( z;a\right) \rightarrow \alpha {\cal P%
}\left( z;a\right)
\end{equation}
In this case will be valid the Legendre transformation between the
fundamental potentials of the ensembles:

\begin{equation}
S_{B}\left( I;a\right) \simeq 
\mathrel{\mathop{\min }\limits_{\beta }}%
\left\{ \beta \cdot I-{\cal P}\left( \beta ;a\right) \right\}
\end{equation}

When the minimum requirement is not satisfied it is due to the occurrence of
critical phenomena in the systems i.e., phase transitions. Outside the
thermodynamic limit or during the phase transitions the only way to describe
the system is microcanonically. Ordinarily we have to deal with finite
systems, so that, we need to precise when we could consider that the
thermodynamic limit takes place in it.

\subsection{The model}

Let us consider an isolated container with volume $V$ in which the
electromagnetic radiation has been confined and it can not scape. Let us
suppose too that this container is the sufficiently large in order to be
valid the continuum approximation for the occupational states density:

\begin{equation}
g_{p}\left( V\right) =2\frac{V4\pi p^{2}dp}{\left( 2\pi \hslash \right) ^{3}}%
=\frac{V}{\pi ^{2}\hslash ^{3}}p^{2}dp
\end{equation}

The only one integral of movement that we consider here is the total energy, 
$E$. For the state with momentum $p$ the correspondent energy eingenvalue is 
$\varepsilon _{p}=pc$, where $c$ is the speed of light in vacuum. The Planck%
\'{}%
s potential in this case is given by:

\begin{eqnarray}
{\cal P}\left( z;V\right) &=&\frac{V}{\pi ^{2}\hslash ^{3}}\int_{0}^{+\infty
}p^{2}\ln \left[ 1-\exp \left( -zpc\right) \right] dp \\
&=&-\frac{V}{3\pi ^{2}\hslash ^{3}}zc\int_{0}^{+\infty }\frac{p^{3}dp}{\exp
\left( zpc\right) -1}  \nonumber \\
&=&-\frac{\widetilde{\sigma }V}{3}\frac{1}{z^{3}}\text{ \ \ \ \ \ }\left( 
\text{with \ }\widetilde{\sigma }=\frac{\pi ^{2}}{15\hslash ^{3}c^{3}}\right)
\nonumber
\end{eqnarray}

From the above relation is inferred that the thermodynamic limit is carried
out when $\left\{ V\rightarrow \infty \text{ }\left| \frac{E}{V}\sim
const\right. \right\} $. Substituting $z$ by $\beta $, in correspondence
with the canonical description, the total energy of the system is given by:

\begin{equation}
E=\frac{\partial }{\partial \beta }{\cal P}\left( \beta ;V\right) =%
\widetilde{\sigma }V\beta ^{-4}
\end{equation}
arriving this way to the {\em Stephan-Boltzmann%
\'{}%
s law}:

\begin{equation}
\rho =\frac{E}{V}=\frac{4}{c}\sigma T^{4}
\end{equation}
where $\sigma =\frac{\pi ^{2}k^{4}}{60\hslash ^{3}c^{2}}$ is the
Stephan-Boltzmann constant. On other hand, the pressure is expressed as: 
\begin{equation}
p=-kT\frac{\partial }{\partial V}{\cal P}\left( \beta ;V\right) =\frac{1}{3}%
\widetilde{\sigma }V\beta ^{-4}\equiv \frac{1}{3}\rho
\end{equation}
Applying the Legendre transformation we can find the information entropy of
the system, $S_{c}(E;V)$:

\begin{equation}
S_{c}(E;V)=\beta E-{\cal P}\left( \beta ;V\right) =\frac{4}{3}\widetilde{%
\sigma }V\beta ^{-3}
\end{equation}

\begin{equation}
S_{c}(E;V)=\frac{4}{3}\left( \widetilde{\sigma }VE^{3}\right) ^{\frac{1}{4}}
\label{ce}
\end{equation}

According to the Eq.[\ref{id}] the states density is obtained by the
calculation of the following integral:

\begin{equation}
\Omega \left( E;V\right) =\int_{-\infty }^{+\infty }d\widetilde{\xi }\exp
\left( i\overline{\xi }E+\frac{1}{3}\widetilde{\sigma }V\frac{1}{\left( i%
\overline{\xi }\right) ^{3}}\right)
\end{equation}
whose integration yields:

\begin{equation}
\Omega \left( E;V\right) =\frac{1}{E}\stackrel{\infty }{%
\mathrel{\mathop{\sum }\limits_{n=1}}%
}\frac{1}{n!\left( 3n-1\right) !}\left( \frac{1}{3}\widetilde{\sigma }%
VE^{3}\right) ^{n}
\end{equation}

The function $W\left( E;V\right) $ in the definition of the Boltzmann%
\'{}%
s entropy (the Eq.[\ref{bol}]) is obtained multiplying the states density by
a suitable energy constant, $\varepsilon _{0}$. A reasonable choice of $%
\varepsilon _{0}$ for the blackbody radiation is:

\begin{equation}
\varepsilon _{0}=\frac{\pi \hslash c}{V^{\frac{1}{3}}}=p_{0}c
\end{equation}
where $p_{0}$ is the length of the elemental cells in the momentum space for
the present problem. This way, the function $W\left( E;V\right) $ is given
by:

\begin{equation}
W\left( E;V\right) =\left( \frac{256\pi ^{5}}{1215}\right) ^{\frac{1}{3}}%
\frac{1}{\left( S_{c}\right) ^{\frac{4}{3}}}H\left( S_{c}\right) 
\end{equation}
where $S_{c}$ is the canonical entropy of the Eq.[\ref{ce}] and the
function\ $H\left( x\right) $ is given by the expression:

\begin{equation}
H\left( x\right) =\stackrel{\infty }{%
\mathrel{\mathop{\sum }\limits_{n=1}}%
}\frac{1}{n!\left( 3n-1\right) !}\left( 3^{3}\left( \frac{x}{4}\right)
^{4}\right) ^{n}  \label{sh}
\end{equation}
and therefore, the Boltzmann%
\'{}%
s entropy is expressed by:

\begin{equation}
S_{B}\left( E;V\right) =\ln H\left( S_{c}\left( E;V\right) \right) -\frac{4}{%
3}\ln S_{c}\left( E;V\right) +const
\end{equation}

In order to analyze the realization of the thermodynamic limit let us to
find the microcanonical temperature, $T_{m}$:

\begin{equation}
\beta _{m}=\frac{1}{kT_{m}}=\frac{\partial }{\partial E}S_{B}=\left( \frac{%
\partial S_{B}}{\partial S_{c}}\right) \frac{\partial }{\partial E}S_{c}
\end{equation}
Defining the function $\gamma \left( S_{c}\right) $ by:

\begin{equation}
\gamma \left( S_{c}\right) =\left( \frac{\partial S_{B}}{\partial S_{c}}%
\right)
\end{equation}
we can express the microcanonical temperature as:

\begin{equation}
T_{m}=\gamma ^{-1}\left( S_{c}\right) T_{c}=\gamma ^{-1}\left( S_{c}\left(
E;V\right) \right) \left( \frac{c}{4}\frac{E}{\sigma V}\right) ^{\frac{1}{4}}
\label{ebm}
\end{equation}

The Eq.[\ref{ebm}] constitutes the microcanonical generalization of the
Stephan-Boltzmann law (see fig.1). The function $\gamma ^{-1}\left(
S_{c}\right) $ allows us to evaluate the convergency of the microcanonical
temperature, $T_{m}$, to the canonical one, $T_{c}$ (see fig.2). In the
thermodynamic limit we expect that it approaches to the unity. This means
that the function $H\left( x\right) $ should have an exponential behavior
for $x\gg 1$. In effect, for large values of $x$ the {\em n-th term} in the
serie Eq.[\ref{sh}] is comparable with:

\begin{equation}
\frac{1}{n!\left( 3n-1\right) !}\left( 3^{3}\left( \frac{x}{4}\right)
^{4}\right) ^{n}\sim \frac{1}{\left( 4n\right) ^{4n}}x^{4n}\sim \frac{x^{4n}%
}{\left( 4n\right) !}
\end{equation}
and therefore, $H\left( x\right) \propto \exp \left( x\right) $. A better
analysis shows that $H\left( x\right) $ converges asymptotically for large
values of $x$ to:

\begin{equation}
H\left( x\right) \simeq Cx^{\theta }\exp \left( x\right)
\end{equation}
where $C=0.171(2)$ and $\theta =0.501(4)$ (this asymptotical behavior is
almost exact with great accuracy for $x>80$). From here, we find the
asymptotic behavior for the Boltzmann%
\'{}%
s entropy:

\begin{equation}
S_{B}\simeq S_{c}-\left( \frac{4}{3}-\theta \right) \ln S_{c}
\end{equation}
and the function $\gamma \left( S_{c}\right) $:

\begin{equation}
\gamma \left( S_{c}\right) \simeq 1-\left( \frac{4}{3}-\theta \right) \frac{1%
}{S_{c}}
\end{equation}

We can say with a precision of $10^{-3}$ that the thermodynamic limit is
established in the system when $S_{c}>S^{\ast }=800$.\ On other hand, the
total energy of the system must satisfy the exigency:

\begin{equation}
E>\left( \frac{1215}{256\pi ^{5}}\left( S^{\ast }\right) ^{4}\right) ^{\frac{%
1}{3}}\epsilon _{0}\approx 1.8\times 10^{3}\epsilon _{0}
\end{equation}

This numerical result shows the high degree of accuracy that the traditional
methodology offers to the macroscopic description of the blackbody radiation
in ordinary conditions. In terms of the temperature the above condition is
rewritten as:

\begin{equation}
kT_{c}>\left( \frac{45}{4\pi ^{5}}S^{\ast }\right) ^{\frac{1}{3}}\epsilon
_{0}\approx 3.1\epsilon _{0}
\end{equation}

It is very interesting to evaluate numerically all the above conditions. For
the elemental energy constant, $\epsilon _{0}$, we have:

\begin{equation}
\epsilon _{0}=\allowbreak \frac{6.\,\allowbreak 240\,6\times 10^{-25}}{V^{%
\frac{1}{3}}}%
\mathop{\rm J}%
\text{ (}V\text{ in }m^{3}\text{)}
\end{equation}
and therefore:

\begin{equation}
T_{c}>\frac{\allowbreak 0.\,\allowbreak 139\,52}{V^{\frac{1}{3}}}%
\mathop{\rm K}%
\text{ (}V\text{ in }m^{3}\text{)}  \label{ct2}
\end{equation}

This is a very important result because it shows that the predictions of our
model could be corroborated in low temperature experiments. Let us now to
study the microcanonical modifications to the Planck%
\'{}%
s formula for the spectral density of energy. From the definition Eq.[\ref
{do}] we can obtain the particles density by mean of the integral:

\begin{equation}
\Omega \left( E;V\right) n_{p}\left( E;V\right) =\int_{-\infty }^{+\infty }d%
\widetilde{\xi }\exp \left( i\overline{\xi }E+\frac{1}{3}\widetilde{\sigma }V%
\frac{1}{\left( i\overline{\xi }\right) ^{3}}\right) \frac{g_{p}\left(
V\right) }{\exp \left( i\overline{\xi }\varepsilon _{p}\right) -1}
\end{equation}
Easily is obtained the following result:

\begin{equation}
n_{p}\left( E;V\right) =g_{p}\left( V\right) \stackrel{N_{p}}{%
\mathrel{\mathop{\sum }\limits_{n=1}}%
}\frac{W\left( E-n\varepsilon _{p};V\right) }{W\left( E;V\right) }\text{ \ \
with }N_{p}=\left[ \frac{E}{\varepsilon _{p}}\right]
\end{equation}

Multiplying the particles density by the energy eingenvalue of the state,
rewriting it again in terms of the frequencies making use of the Planck%
\'{}%
s relation:

\begin{equation}
\varepsilon _{p}=pc=\hslash \omega
\end{equation}
we have finally the microcanonical spectral energy density, $u_{m}\left(
\omega ;E,V\right) $:

\begin{equation}
u_{m}\left( \omega ;E,V\right) d\omega =\frac{\omega ^{2}}{\pi ^{2}c^{3}}%
\hslash \omega \stackrel{N_{\omega }}{%
\mathrel{\mathop{\sum }\limits_{n=1}}%
}\frac{W\left( E-n\hslash \omega ;V\right) }{W\left( E;V\right) }d\omega 
\text{ \ \ with }N_{\omega }=\left[ \frac{E}{\hslash \omega }\right]
\label{pe}
\end{equation}

Among the most interesting peculiarities of the above expression we found a 
{\em cut off at the frequency} $\omega _{c}$:

\begin{equation}
\omega _{c}=\frac{E}{\hslash }  \label{fc}
\end{equation}
around the same one the function $u_{m}\left( \omega ;E,V\right) $ behaves
as:

\begin{equation}
u_{m}\left( \omega \approx \omega _{c};E,V\right) d\omega \simeq \frac{1}{%
6\pi ^{2}c^{3}}\frac{\widetilde{\sigma }VE^{3}}{W\left( E;V\right) }\left(
\omega _{c}-\omega \right) ^{2}d\omega
\end{equation}

On other hand, at the {\em low frequency limit} we have:

\begin{equation}
u_{m}\left( \omega \approx 0;E,V\right) d\omega \simeq \frac{\omega ^{2}}{%
\pi ^{2}c^{3}}kT_{m}d\omega  \label{rj}
\end{equation}
This is the microcanonical generalization of the Raleigh-Jean%
\'{}%
s formula, where $T_{m}$ is the microcanonical temperature. Finally, during
the thermodynamic limit:

\begin{equation}
u_{c}\left( \omega ;E,V\right) d\omega =\frac{\omega ^{2}}{\pi ^{2}c^{3}}%
\hslash \omega 
\mathrel{\mathop{\lim }\limits_{V\rightarrow \infty ,\frac{E}{V}\sim const}}%
\left( \stackrel{N_{\omega }}{%
\mathrel{\mathop{\sum }\limits_{n=1}}%
}\exp \left[ \ln W\left( E-n\hslash \omega ;V\right) -\ln W\left( E;V\right) %
\right] \right) d\omega
\end{equation}

\begin{equation}
u_{c}\left( \omega ;T_{c}\right) d\omega =\frac{\omega ^{2}}{\pi ^{2}c^{3}}%
\frac{\hslash \omega d\omega }{\exp \left( \beta _{c}\hslash \omega \right)
-1}
\end{equation}
we arrive to the celebrated {\em Planck%
\'{}%
s formula}.

\section{conclusions}

In the present work we have found the conditions in which can be considered
that the thermodynamic limit has been established for the electromagnetic
radiation confined in an isolated container with volume $V$. Our
calculations showed that the total energy, $E$, only needs to be some
thousands of times the elementary energy constant, $\varepsilon _{0}$. In
normal conditions the traditional description of the blackbody radiation is
extraordinary exact. This fact clarifies the sterility of the intents of
finding nonextensive effects under these conditions. This conclusion is
supported by direct experiments trying to find nonextensive effects in the
cosmic microwave background radiation from the data obtained via Cosmic
Background Explorer Satellite by Mather et al\cite{cbes}. The analysis of
these data through the Statistic of Tsallis showed the great accuracy of the
traditional description: with 95\% of confidence $\left| q-1\right|
<3.6\times 10^{-5}$(the Tsallis%
\'{}%
s Thermodynamic becomes in the traditional thermodynamic at $q=1$).

However, for the blackbody radiation confined at low temperatures in a
finite and isolated container, our results predicts desviations with respect
to the traditional analysis if those temperatures do not satisfy the
condition in Eq.[\ref{ct2}]. This nonextensive effects could be detected
through of measurements of the spectral density of energy at low energies.
According to the Eq.[\ref{pe}], at high frequencies must be detected the cut
off frequency $\omega _{c}$ which is related with the total energy via the
Eq.[\ref{fc}]. The microcanonical temperature could be found analyzing the
low frequency region of the spectral density by mean of the Eq.[\ref{rj}].
This way we can check the validity of the Stephan-Boltzmann%
\'{}%
s law. Unfortunately there is no a ideal containers, since they are composed
by a big number of particles, that is, they are extensive systems. This fact
leads to hard difficulties for the practical realization of those
experimental measurements.

\end{document}